\documentclass{ws-p8-50x6-00}
\newcommand{\mum}{$\,\mu$m}

\begin{document}

\title{Known and unknown SCUBA sources}

\author{Douglas Scott,}
\author{Colin Borys, Mark Halpern, Anna Sajina,}
\address{Department of Physics \& Astronomy,
         University of British Columbia, 
         Vancouver, BC V6T 1Z1\ \ CANADA\\E-mail: dscott@astro.ubc.ca}

\author{Scott Chapman,}
\address{Observatories of the Carnegie Institution of Washington,
         Pasadena, CA 91101\ \ USA}

\author{and Greg Fahlman}
\address{Canada-France-Hawaii Telescope,
         Kamuela, Hawaii 96743\ \ USA}

\maketitle

\abstracts{Summary and discussion of some projects to use SCUBA to target
sources selected at other wavebands, as well as to find new sub-mm
galaxies in `blank fields'.
}

\section{Introduction}
The sub-mm waveband has opened up for cosmology, and through this window
we can hope to glimpse some answers to a number of related puzzles:
What are the brightest sub-mm galaxies?
What sorts of galaxies make up the Far-IR Background (FIB)?
When did the Universe form the bulk of its stars?
How important is dust obscuration for obtaining a full star-formation
census?

SCUBA\cite{scuba}
has been instrumental in establishing this new field of sub-mm
high redshift astronomy.  This meeting has been dominated by SCUBA-based
surveys, and stands as a testament to the instrument builders who produced
the best camera of its kind at just the right time.

\begin{figure}[ht]
\epsfxsize=30pc
\center\epsfbox{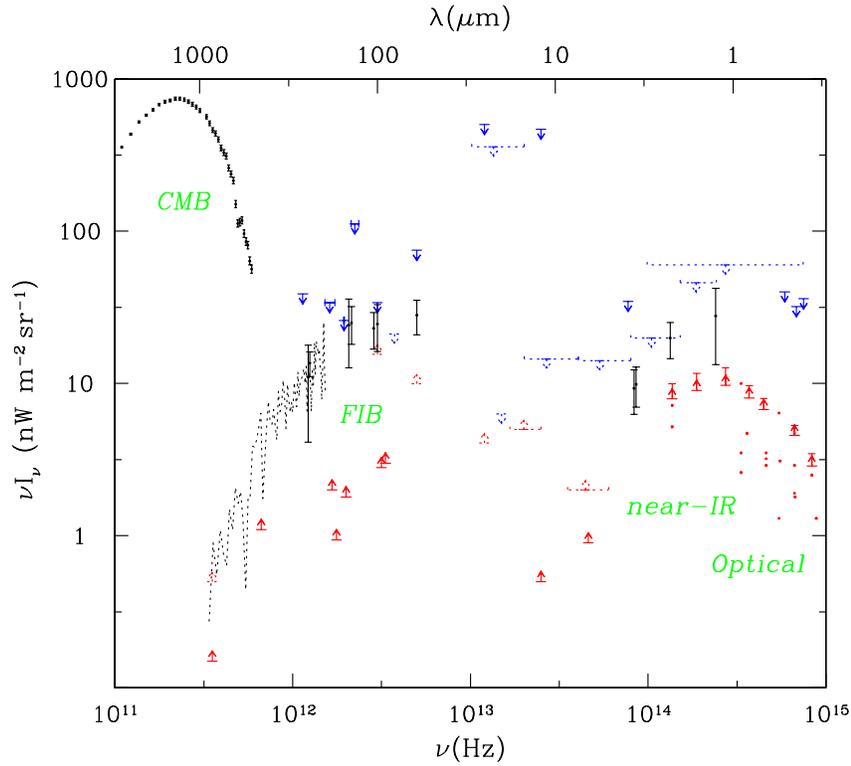} 
\label{fig:firb}
\caption{Estimates for the cosmic background in the IR and neighbouring
wavebands.  This is a blown-up and up-dated plot of Figure~26.2 in
{\it Astrophysical Quantities}.
Lower limits generally come from the integrated brightness of resolved sources.
Limits which are more model-dependent
are indicated by dotted lines.  The FIB has been slowly emerging from
careful analyses of (largely) {\sl COBE\/} data.  It appears to contain more
energy than the optical background (usually assumed to be the sum of the
deepest HST counts).  It also appears to be broader than the blackbody
CMB peak, possibly indicating that it comes from a range of redshifts.
References for the points are given in the text.}
\end{figure}

Many basic questions have already been answered by the SCUBA data which
are in hand, and a coherent picture has developed -- typical
SCUBA sources are the $z\,{\sim}\,$2--3 counterparts of locally well-studied
Ultraluminous Infrared Galaxies, and these are
considerably more common at an epoch which was perhaps important for the
formation of elliptical galaxies.
However, there are still a great many details to understand about the
exact composition of the SCUBA-bright sources, including how they overlap with
populations selected at other wavelengths, how the ${\sim}\,$mJy sources
might differ from the brighter ones, what the clustering strength of the
SCUBA sources might be, and what fraction could be
at still higher redshifts.
Here we will avoid reviewing all of the different
studies (many of which are covered in other
presentations), but will concentrate on projects carried out by our own group.

We have been undertaking a set of related studies, focussing on either
blank fields to detect sub-mm sources, or on characterizing the sub-mm
properties of objects selected at other wavelengths.

\section{Known sources}
\subsection{FIRBACK}

A summary of current information on extragalctic background radiation is
presented in Fig.~1.  In the far-IR region, the important detections are
derived from DIRBE data at 240\mum, 140\mum, 100\mum\ and
60\mum\cite{Hauser,Lag00,FDS00} and FIRAS data at longer
wavelengths\cite{Fixsen,Lag99}.  Detections in the near-IR have also been
derived from DIRBE data\cite{DwekA,GorjianWC,WrightR,Wright}.  
Upper limits come from DIRBE\cite{Hauser}, various rocket
experiments\cite{Kawada,Matsuura},
and indirectly from $\gamma-$rays\cite{Biller,Guy}.
Optical lower limits (which in some cases have probably converged
to near the background values) come from galaxy counts\cite{Leinert,MadauP}.
Lower limits throughout the IR come from source counts obtained by
{\sl ISO}\cite{Altieri,Metcalfe,Elbaz,Clements,Matsuhara,Juvela},
SCUBA\cite{Hughes,Blain,Blain450} and
{\sl IRAS\/}\cite{HackingS,Gregorich} -- some of these values are also
model-dependent.

Now that a cosmic FIB is emerging from the data, and that it makes a
substantial contribution to the energy budget, the obvious question is:
what are the galaxies which make up the FIB?
This is the main motivation behind the FIRBACK project, which 
made deep $170\,\mu$m images of the sky with ISO, to break up the
background into the sources which comprise it\cite{FIRB1,FIRB2,FIRB3,Lag00}.
Data from the ISOPHOT instrument is the only currently available information
for studying the composition of the FIB {\it which comes directly from
wavelengths where the background peaks}\cite{AQ}.
We have been studying the SCUBA properties of some of these already known
FIRBACK-selected sources.
Because of the large ISOPHOT beam, this is only really feasible for sources
with reliable radio identifications.  Photometry with SCUBA allows you to go
deeper than mapping in the same integration time.  We have also successfully
been using `3-bolometer chopping', by choosing the chop position to be
approximately coincident with two bolometers in the inner ring of the array,
so that those negative signals can be added in, to improve the
signal-to-noise\cite{LBGs}.

The combination of ISOPHOT + VLA + SCUBA selects galaxies covering a range
of redshifts but peaking at $z\,{\sim}\,1$ (supported by the handful of
objects for which redshifts have already been obtained).
The study of FIRBACK galaxies at $z\,{\sim}\,1$ thus
bridges the gap between the well-studied $z\,{\sim}\,0$ ULIRGs and the
$z\,{\sim}\,2$--3 SCUBA sources.  At these somewhat more modest redshifts the
SEDs can be realistically constructed over a wide range of
wavelengths.

\begin{figure}[ht]
\epsfxsize=25pc 
\epsfbox{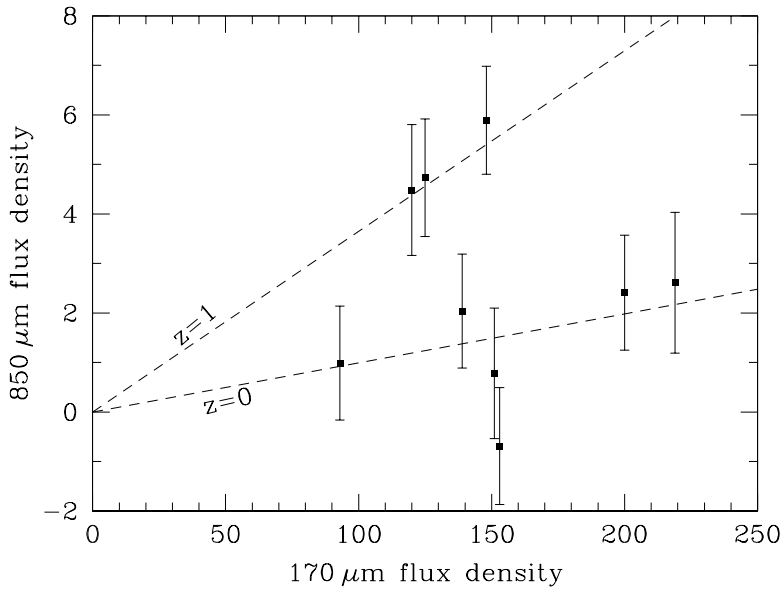} 
\label{fig:z850170}
\caption{Measured 170\mum\ vs 850\mum\ flux densities
for our first sample of FIRBACK sources.  The two dashed lines
are $T_{\rm d}=50\,$K, $\beta=1.5$ modified blackbodies at $z=0$ and $z=1$.
This shows that just the combination of 850\mum\ and 170\mum\ can
give crude redshift information about these galaxies (three seem to lie
at higher $z$ than the others) -- more detailed SED fitting of the full
radio/submm/far-IR/mid-IR/near-IR/optical data will reveal much more of
course.
}
\end{figure}

The total number of sources detected in the ELAIS `N1' field\cite{ELAIS}
is ${\sim}\,120$ with $S_{170}\,{>}\,120\,$mJy.  They
comprise about 10\% of the cosmic Far-IR
Background at this wavelength, and an entirely unknown fraction of the
longer wavelength sub-mm background.  Note that
SCUBA `blank fields' only tell you about the background at 850$\,\mu$m, where
$\nu I_\nu\,{\sim}\,30$ times lower than its peak value (see Fig.~1).
Our first set of results from a sample of 10 objects in the N1 field
was presented in Scott et al.~(2000)\cite{Scott}.
There we firmly detected 4 sources at 850$\,\mu$m.  Statistically the
sample was detected at $7.5\sigma$ at 850$\,\mu$m and $4\sigma$ even at
450$\,\mu$m.  Crude photometric redshifts can be obtained simply by
comparing the 170$\,\mu$m and 850$\,\mu$m fluxes, as is shown in
Fig.~2.

Joint fits to $L_{\rm FIR}$ and $z$ show that 3 of the galaxies are
consistent with luminosities typical of Arp$\,220$, but at $z\,{\sim}\,1$.
The others seem more likely to be at redshifts intermediate between 0 and
1, and intrinsically fainter (but still representing a population that
hardly exists locally).
Multi-wavelength studies are underway (Lagache et al.~in preparation), and
the SCUBA work should continue, with a sample covering a wider range of
ISO fluxes and properties at other wavelengths.

\subsection{The Blob}

Extensive spectroscopic surveys have shown strong clustering among
the star-forming Lyman Break Galaxy population at $z\,{\simeq}\,3$
(which is discussed by Chapman et al.~in these proceedings).
A high contrast overdensity of LBGs at $z\,{=}\,3.09$ in the SSA22 region
was discovered by using deep narrow-band Ly$\,\alpha$ imaging to identify
at least 160 members of the structure.  This region appears to be about
6 times more overdense in LBGs than the general field, and the region
centred on the most extended  Ly$\,\alpha$ emission is overdense by
roughly a further factor of 2.  This is where we centred our SCUBA
map\cite{SCUBAblob}.
The brightest source in that map (Fig.~3) is ${\simeq}\,20\,$mJy,
and centred precisely on the extended Ly$\,\alpha$ region referred to
as `Blob~1' in Steidel et al.~(2000)\cite{blob}.

\begin{figure}[ht]
\epsfxsize=25pc
\epsfbox{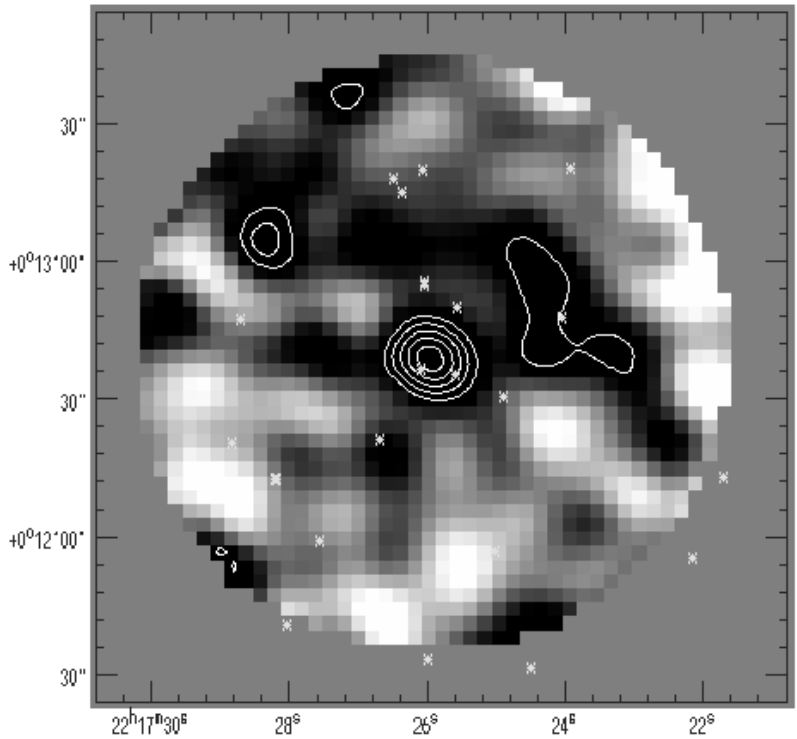}
\label{fig:blob}
\caption{SCUBA 850\mum\ map centred on `Blob~1' of the Ly$\,\alpha$
survey of Steidel et al.~(2000).  Crosses mark the positions of
known $z\,{\simeq}\,3$ galaxies.  Black is positive emission here, and
the contours are at 3, 4, 5, 6 and $7\sigma$.
}
\end{figure}

Such sub-mm emission is unexpected, since dust usually implies no
Ly$\,\alpha$.  The simplest explanation is an obscured AGN, which provides
the excitation for the extended Ly$\,\alpha$, and whose nucleus is responsible
for the sub-mm emission.  It is also possible that some of the signal could
be from the Sunyaev-Zel'dovich increment in a very overdense group, say,
forming at $z\,{\sim}\,3$, but conditions would have to be rather extreme for
this to be the case.  Nevertheless it is worth following up in detail at other
wavelengths to check.  This surprising result suggests that
the combination of SCUBA and narrow-band Ly$\,\alpha$ studies
could prove to be a fruitful one.

There are some additional sources in our SCUBA map, indicating
an overdensity in sub-mm sources, matching that in the LBG population.
It would be interesting to map this region more extensively to determine
if it is a region of enhanced clustering of SCUBA sources.

\section{Unknown sources}

We have made reasonably large maps of the Hubble Deep Field
Flanking Fields and part of the Groth Strip (see the contribution by
Borys et al.) in order to find the brightest SCUBA sources in those
well-studied regions.  In addition  we have a large amount of `blank sky'
data from the off-centre bolometers obtained during photometry observations
of known sources.  Although too under-sampled to be much use for mapping,
these data can nevertheless be studied statistically\cite{BorysCS}.

\subsection{Cluster survey}

We have also looked at several rich cluster fields.
Here the lensing amplification boosts the number counts (as pioneered
by Blain, Ivison, Smail and collaborators).
We have collected data on 9 separate clusters:
Cl\,0016+16, MS\,0451-03, Abell\,520, Zwicky\,3146, MS\,1054-03, MS\,1455+22,
Abell\,2163, Abell\,2219 and Abell\,2261.  The clusters were mapped to
a variety of depths under different observing conditions.  But the
resulting maps typically contain a couple of convincing sources.  Details
are presented in Chapman et al.~(2000)\cite{Clusters}.

\begin{figure}[ht]
\epsfxsize=25pc 
\epsfbox{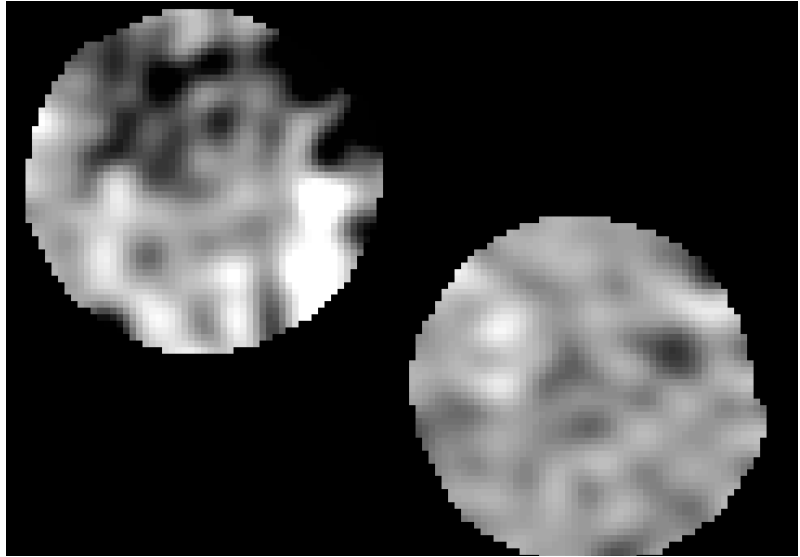} 
\label{fig:a520}
\caption{SCUBA 850\mum\ maps of Abell$\,520$, at two different positions
which almost overlap.  Here positive emission is white and the noisy edge
regions have been clipped.
The map at upper left is from our own data and shows two very bright sources
at ${\sim}\,15\,$mJy and ${\sim}\,30\,$mJy.  The lower right map is from the
SCUBA archive, is less noisy, and shows two ${\sim}\,10\,$mJy sources.}
\end{figure}

Instead of running through each cluster field in detail, let us focus
on only one, Abell\,520.  Fig.~4 shows our map (upper left), together with an
almost overlapping map taken from the SCUBA archive.  Each map appears to
show 2 detections.  There are no obvious near-IR or radio identifications for
these sources, but the follow-up data are, as yet, not very deep.
One of the sources we discovered (SMMJ$\,04543{+}0257$) appears to be
${\simeq}\,30$mJy, placing it among the brightest `blank sky' sources.

\section{Conclusions}

Further clues to the nature of sub-mm sources and how they overlap with
optical-, IR- and radio-selected galaxies will be found through the joint
approach of investigating the sub-mm properties of already known objects,
and finding identifications for possibly unknown objects which are bright
in the sub-mm.  Studies of the clustering of SCUBA sources are just in their
infancy, but should provide further cosmological information.  The crucial
thing is to identify the SCUBA sources at other wavelengths -- this has
proved to be difficult in practice, and the sources for which
is {\it most\/} difficult are potentially the most interesting!

\section*{Acknowledgments}
DS wishes thanks all his collaborators on these observational projects,
which keep him away from his day job.

\end{document}